\documentclass{article}

\usepackage{arxiv}

\usepackage[utf8]{inputenc} 
\usepackage[T1]{fontenc}    
\usepackage{hyperref}       
\usepackage{url}            
\usepackage{booktabs}       
\usepackage{amsfonts}       
\usepackage{nicefrac}       
\usepackage{microtype}      
\usepackage{lipsum}		
\usepackage{amsmath,graphicx}
\usepackage{natbib}
\usepackage{doi}
\usepackage{float}
\usepackage{authblk}

\title{LESEN: Label-Efficient deep learning for Multi-parametric MRI-based Visual Pathway Segmentation}

\author[1, 2]{Alou Diakite}
\author[1]{Cheng Li}
\author[3]{Lei Xie}
\author[3]{Yuanjing Feng}
\author[1, 2]{Hua Han}
\author[1, 4]{Shanshan Wang \thanks{ss.wang@siat.ac.cn}}
\affil[1]{Paul C. Lauterbur Research Center for Biomedical Imaging, Shenzhen Institute of Advanced Technology, Chinese Academy of Sciences, Shenzhen, China}
\affil[2]{University of Chinese Academy of Sciences, Beijing, China}
\affil[3]{Zhejiang University of Technology, Hangzhou, China}
\affil[4]{Peng Cheng Laboratory, Shenzhen, China}

\date{}

\renewcommand{\shorttitle}{\textit{arXiv} Template}

\hypersetup{
pdftitle={A template for the arxiv style},
pdfsubject={q-bio.NC, q-bio.QM},
pdfauthor={David S.~Hippocampus, Elias D.~Striatum},
pdfkeywords={First keyword, Second keyword, More},
}

\begin{document}
\maketitle

\begin{abstract}
Recent research has shown the potential of deep learning in multi-parametric MRI-based visual pathway (VP) segmentation. However, obtaining labeled data for training is laborious and time-consuming. Therefore, it is crucial to develop effective algorithms in situations with limited labeled samples. In this work, we propose a label-efficient deep learning method with self-ensembling (LESEN). LESEN incorporates supervised and unsupervised losses, enabling the student and teacher models to mutually learn from each other, forming a self-ensembling mean teacher framework. Additionally, we introduce a reliable unlabeled sample selection (RUSS) mechanism to further enhance LESEN's effectiveness. Our experiments on the human connectome project (HCP) dataset demonstrate the superior performance of our method when compared to state-of-the-art techniques, advancing multimodal VP segmentation for comprehensive analysis in clinical and research settings. The implementation code will be available at: https://github.com/aldiak/Semi-Supervised-Multimodal-Visual-Pathway- Delineation.
\end{abstract}


\section{Introduction}
In recent years, multi-parametric MRI has emerged as a powerful tool in the field of visual pathway (VP) segmentation \cite{Li2021}. By combining different imaging sequences, such as T1-weighted (T1-w) and fractional anisotropy (FA), researchers have been able to overcome the limitations of individual imaging sequences and gain a more comprehensive understanding of the VP. This can be a powerful tool in the diagnosis and treatment of various diseases affecting the visual system, including optic neuritis and optic nerve glioma.

During the past years, deep learning (DL) approaches, such as convolutional neural networks (CNNs) and their variants, have shown significant advancements in multi-parametric MRI-based VP segmentation \cite{Li2021, Mansoor2016, Zhao2019, Xie2023_1}. These methods can automatically learn hierarchical representations from the multi-parametric MR images and effectively capture the complementary characteristics.

However, one of the major challenges in these DL-based methods is the laborious and time-consuming process of obtaining labeled training data \cite{Bai2023, yu2019uncertainty}. Manual annotation is prone to errors and requires significant expertise, making it impractical for large-scale studies. Therefore, there is a critical need to develop algorithms that can achieve accurate segmentation performance even in situations with limited labeled samples.

In this work, we propose a label-efficient self-ensembling network (LESEN) to address the issue of limited labeled training data in VP segmentation. The LESEN framework consists of two subnetworks: a student model and a teacher model . By incorporating both supervised loss from labeled data and unsupervised loss from unlabeled data, the student and teacher models can mutually learn from each other, leveraging a self-ensembling mean teacher mechanism. This approach allows us to effectively utilize the available labeled data while also benefiting from the information contained in the unlabeled data. Additionally, we introduce a reliable unlabeled sample selection (RUSS) mechanism to further enhance the effectiveness of LESEN. RUSS aims to identify the most informative and reliable samples from the unlabeled dataset, further improving the segmentation performance.

To evaluate the performance of our proposed method, experiments were conducted using the open-source human connectome project (HCP) dataset. The results demonstrate the superior performance of LESEN compared to state-of-the-art techniques in multi-parametric MRI-based VP segmentation. Our approach not only enhances the accuracy and efficiency of VP segmentation but also contributes to comprehensive analysis in clinical and research settings.


\begin{figure}[htb]

\begin{minipage}[b]{1.\linewidth}
  \centering
  \centerline{\includegraphics[width=14.cm]{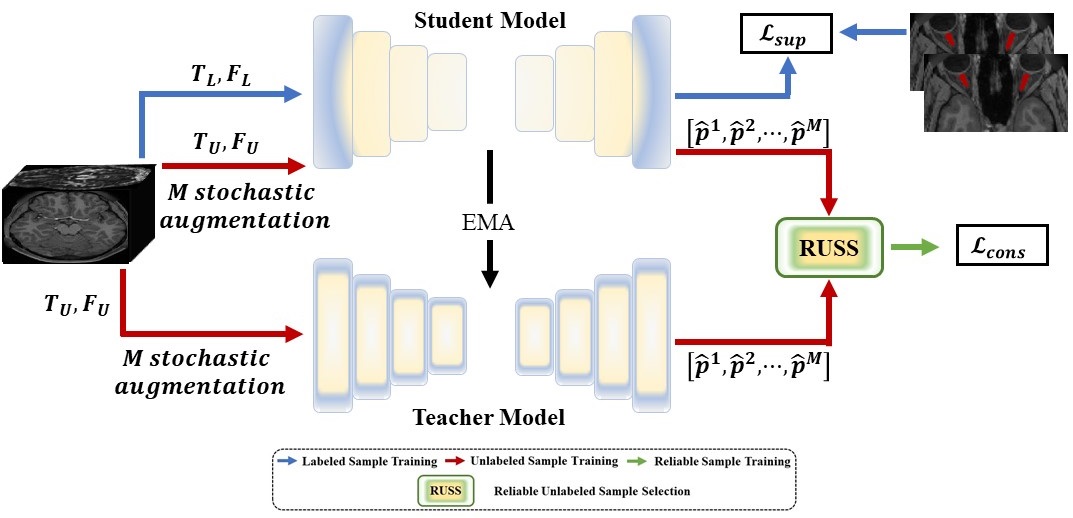}}
\end{minipage}
\caption{The overall architecture of our proposed method. There are two subnetworks, including a student model and a teacher model with each model taking two inputs. The models share the same structure and are used to compute both supervised and unsupervised consistency losses during training.}
\label{fig1}
\end{figure}

\section{Methodology}
Let T and F be the T1-weighted (T1w) and fractional anisotropy (FA) images, respectively. We denote the labeled datasets of T1w modality and FA modality as $T_L = {(x_{i},y_{ij})}$ and use $F_L = {(x_{j},y_{ij})}$ , respectively. Similarly, the unlabeled images in the T1w and FA sequences are denoted as $T_U = {(x_{u(i)})}$ and $F_U = {(x_{u(j)})}$, respectively. The architecture of the proposed method is illustrated in \textbf{Figure \ref{fig1}}, which consists of a self-ensembling mean teacher (SE-MT) with a reliable unlabeled sample selection (RUSS) mechanism to select reliable unlabeled samples. We adopted the standard U-Net \cite{Ronneberger2015} shaped encoder-decoder structure as the backbone for segmentation with spatial attention blocks to selectively attend to relevant regions or features in each multi-parametric MR image and fuse them.

Given annotated image pairs $(x_{i},y_{ij}) \in T_L$ from T1w images and $(x_{j},y_{ij}) in F_L$ from FA images, we establish a supervised loss function to optimize the parameters of the student model:

\begin{equation} \label{eq_1}
 L_{sup} = - \sum_{n=1}^{k} \mathbf{E}_{[y_{ij}]} \log (\theta^{S}(x_{i}, x_{j})),
\end{equation}

where $\mathbf{E}$ represents an indicator function.

\begin{figure}[htb]

\begin{minipage}[b]{1.\linewidth}
  \centering
  \centerline{\includegraphics[width=9.5cm]{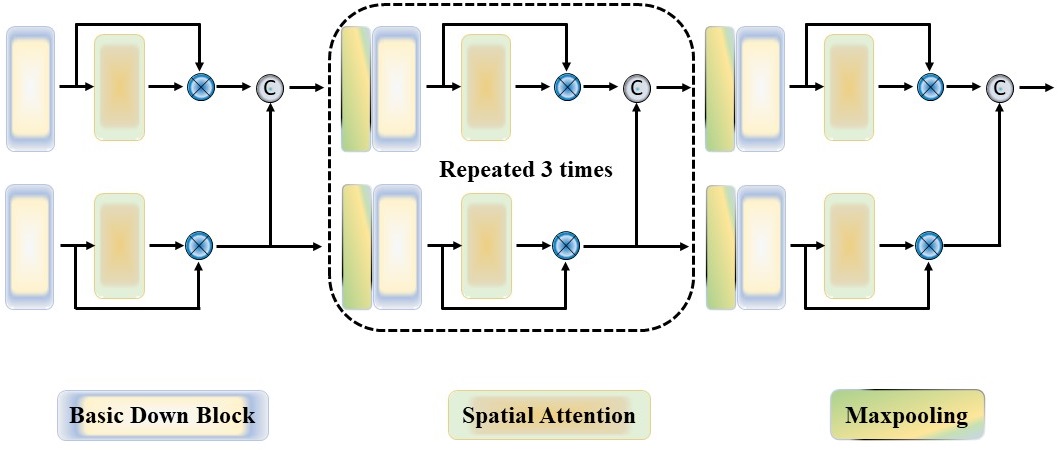}}
\end{minipage}
\caption{The detailed structure of the LESEN's encoder.}
\label{fig2}
\end{figure}

\subsection{Network Architecture}

To effectively fuse the information of different MR imaging sequences, we propose an encoder-decoder network architecture that incorporates spatial attention modules (Figure 2). This module is designed to select and fuse relevant features from T1w and FA images, thereby improving segmentation performance.
Inspired by previous works \cite{Li2019, Xie2023_1}, our baseline network follows a U-shaped encoder-decoder structure. This architecture allows for the incorporation of skip connections between the encoding and decoding layers, ensuring the preservation of spatial information throughout the network. To avoid pixel-wise summation, we use channel-wise concatenation for feature fusion.
To make the network more lightweight, we reduce the convolution kernel number of each layer in the encoder. 
The spatial attention block plays a crucial role in our network architecture. It enables the learning of relevant regions or features of each multi-parametric MR image. In \textbf{Figure \ref{fig2}}, the spatial attention block is depicted, taking features from the T1w and FA sequences streams as input. It produces a weight heatmap that guides the selection of information from both T1w and FA, ensuring that only relevant features are fused.
In the decoder, skip connections are employed to preserve spatial information that may be lost during the decoding process. These connections help to maintain the accuracy of the segmentation results by incorporating high-resolution features from earlier layers.

\subsection{Self-Ensembling Mean Teacher with RUSS}
Although supervised learning allows the constructed encoder-decoder network to learn from labeled training data, the small scale of labeled scans can still hinder the model's performance. To mitigate this problem, we utilize a self-ensembling mean teacher architecture to exploit unlabeled images. Specifically, the teacher model $\theta^{T}$ acts as an Exponential Moving Average (EMA) of the student model $\theta^{S}$. In the $k_{th}$ training step, its parameter is determined as follows:
\begin{equation}  \label{eq_2}
\theta_{k}^{T} = \alpha \theta_{k-1}^{T} + (1 - \alpha) \theta_{k}^{S}
\end{equation}
where $\alpha \in [0, 1]$ is a decay factor controlling the EMA \cite{laine2016temporal}. Then, an unsupervised consistency loss is defined to measure the agreement between the predictions of the student model and the teacher model on unlabeled data. The unsupervised consistency loss is defined as follows:

\begin{equation} \label{eq_3}
 L_{cons} = \sum_{n=1}^{N} \| \theta^{S}(x_{u(i)}, x_{u(j)}) - \theta^{T}(x_{u(i)}, x_{u(j)}) \|^2 
\end{equation}

Equation \ref{eq_3} allows the teacher model to guide the student model to provide more stable and reliable predictions, reducing the impact of noisy or inconsistent samples during training.

To further enhance the framework's performance, we introduced RUSS. 

Given an unlabeled sample $x_{u(i)}$ and $x_{u(j)}$, we perform stochastic augmentations m times to obtain m copies of $x_{u(i)}^{m}$ and $x_{u(j)}^{m}$ under random augmentations (M = 5 in this study). Let $x_{u(i)}^{m}$ and $x_{u(j)}^{m}$  represent the $m_{th}$ augmented sample of $x_{u(i)}$ and $x_{u(j)}$, where $m$ ranges from $1$ to $M$. For each $x_{u(i)}^{m}$ and $x_{u(j)}^{m}$, their corresponding probability vector $\hat{p}^m = \Bigl(\hat{p}_1^{m}, \hat{p}_2^{m}, \ldots, \hat{p}_j^{m}, \ldots, \hat{p}_k^{t}\Bigl)$  can be generated by feeding $x_{u(i)}^{m}$ and $x_{u(j)}^{m}$ into the student and teacher models. Each element $\hat{p}_j^{m}$ of $\hat{p}^m$ represents the probability that $x_{u(i)}^{m}$ and $x_{u(j)}^{m}$ belongs to the $j_{th}$ category. The consistency of the predictions on $x_{u(i)}$ and $x_{u(j)}$ by the student and teacher models can be measured by the standard deviation of the M probability vectors $ [\hat{p}_j^{1}, \hat{p}_j^{2}, \ldots, \hat{p}_j^{M}]$, which is calculated as:

\begin{equation} \label{eq_4}
cons_{\theta^{S}} = - \sum_{j=1}^{k}\sqrt{var\Bigl(\Bigl[\hat{p}_j^{1}, \hat{p}_j^{2}, \ldots, \hat{p}_j^{M}\Bigl]\Bigl)}
\end{equation}

\begin{equation} \label{eq_5}
cons_{\theta^{T}} = - \sum_{j=1}^{k}\sqrt{var\Bigl(\Bigl[\hat{p}_j^{1}, \hat{p}_j^{2}, \ldots, \hat{p}_j^{M}\Bigl]\Bigl)}
\end{equation}

where $cons_{\theta^{S}}$ and $cons_{\theta^{T}}$ denote the consistency of student and teacher models predictions. The negative sign means a larger standard deviation on the probability vector corresponds to a smaller cons value.

In the training phase, each batch of unlabeled samples can be sorted in descending order based on the cons value. Let $L = [l_1, l_2, ..., l_i, ..., l_b]$ be the sorted index list, where $l_i$ represents the index of the sample with the $i_{th}$ largest $cons_{\theta^{S}}$ and $cons_{\theta^{T}}$ in the batch, and $b$ is the batch size of the unlabeled samples.  RUSS of $x_{u(i)}$ and $x_{u(j)}$ for both models is defined as:
\begin{equation} \label{eq_6}
RUSS = 
\begin{cases}
1, & \text{if} \quad i \in [l_1, l_2, \ldots, l_q] \\
0, & \text{if} \quad i \notin [l_1, l_2, ..., l_q]     
\end{cases}
\end{equation}
where $1 \leq q \leq b$ is a threshold that determines how many samples can be selected in the batch. The value of $q$ is determined using a ramp-up function:
\begin{equation} \label{eq_7}
q = int\Bigl(b \times e^{-(1-epoch/total\_epoch)^2}\Bigl)       
\end{equation}
where $epoch$ and $total\_epoch$ represent the current epoch and the total number of epochs, respectively. int(·) denotes the rounding operation.

RUSS is designed based on the intuition that the student and teacher models should provide consistent predictions for $x_{u(i)}$ and $x_{u(j)}$ regardless of the stochastic augmentations. Only samples with the top $q$ largest $cons_{\theta^{S}}$ and $cons_{\theta^{T}}$ in the batch will be retained according to the constraint in Equation \ref{eq_6}. The modified unsupervised consistency loss can be calculated as:

\begin{equation} \label{eq_8}
 L_{cons} = \sum_{n=1}^{N} \| \theta^{S}(x_{u(i)}^{'}, x_{u(j)}^{'}) - \theta^{T}(x_{u(i)}^{'}, x_{u(j)}^{'}) \|^{2} 
\end{equation}
where $x_{u(i)}^{'}$ and $x_{u(j)}^{'}$ represent the unlabeled data that produced consistent prediction for T1 weighted and FA images by using RUSS. $\theta^{S}(\cdot)$ and $\theta^{T}(\cdot)$ represent the mean of the probability vectors of the student and teacher models on $M$ copies of $x_{u(i)}$ and $x_{u(j)}$ under random augmentations.

The combination of the self-ensembling mean teacher model and the RUSS mechanism contributes to the overall effectiveness of our LESEN framework. 

The full objective function for training our model can be formulated as follows:

\begin{equation} \label{eq_9}
\arg \min_{\{\theta^{S}\}} (\alpha L_{sup} + \lambda L_{cons})
\end{equation}

Here, $\alpha = 0.5$ is a hyper-parameter controlling the weight of $L_{sup}$, and $\lambda$ represents the current consistency weight of $L_{cons}$.

\section{Experiments and Results}
\subsection{Dataset and Implementation details}
This study used the Human Connectome Project (HCP) dataset \cite{van2012human} for evaluation. A total of $92$ cases were used, among which $82$ cases were randomly selected for training and the remaining $10$ cases for testing. In our semi-supervised learning setting, $16$ cases of the $82$ training data were labeled and $66$ were unlabeled. During experiments, the data were resized to $128 \times 160 \times 128$. The acquisition parameters and the process of generating the ground truth for this dataset is provided in previous studies \cite{Xie2023_1,Xie2023}. 

During the training phase, we used an initial learning rate of $0.0002$ and a weight decay of $0.00001$, and the model was trained for 200 epochs. The experiments were conducted on a workstation with NVIDIA GeForce RTX  3090 GPUs (24GB). The study addressed the imbalance between foreground and background voxels by adding the Dice Loss to the supervised loss as follows:
\begin{equation} \label{eq_10}
\begin{split}
 L_{sup} = (- \sum_{n=1}^{N} \mathbf{E}_{[y_{ij}]} log (\theta^{S}(x_{i}, x_{j}))\\
 + (1 - \sum_{n=1}^{N} \mathbf{E}_{[y_{ij}]}) [\frac{2\times\theta^{S}(x_{i}, x_{j})}{\|y_{ij}\| + \|\theta^{S}(x_{i}, x_{j})}]
 \end{split}
\end{equation}

\subsection{Results and Analysis}
In this study, we compared our proposed approach with several existing supervised and semi-supervised methods for VP segmentation in order to demonstrate its effectiveness. These methods include Baseline 1, Baseline 2, FuseNet \cite{Xie2023_1}, TPSN \cite{Li2021}, TransAttUnet \cite{chen2023transattunet}, UA-MT \cite{yu2019uncertainty}, and BCP \cite{Bai2023}.  Baseline 1 (T1-weighted images as input) and Baseline 2 (FA images as input) are just standard U-Net trained with single MRI sequence input. For a fair comparison, all comparison models (FuseNet, TPSN, TransAttUnet, UA-MT, and BCP) adopt the same networks as in their original publication. The only modification was made to the input layer if the original model was not built for multi-parametric MRI inputs. For supervised methods, including Baseline 1, Baseline 2, FuseNet, TPSN, and TransAttUnet, were trained using the 16 labeled data. For semi-supervised methods, including, they were trained using the 16 labeled and 66 unlabeled data.

\begin{table}[htbp]
    \caption{Quantitative results of different methods. The best results are bolded}
    \begin{center}
        \begin{tabular}{|c|c|c|}
            \hline
            \textbf{Methods} & \textbf{DSC $\pm$ sdt  (\%)}  & \textbf{ASD $\pm$ std  (mm)} \\                   
            \hline
            Baseline 1       & 71.1$\pm$0.021  & 0.39$\pm$0.10  \\
            \hline
            Baseline 2      & 64.6$\pm$0.07  & 0.48$\pm$0.16 \\
            \hline
            FuseNet         & 73.3$\pm$0.036  & 3.36$\pm$0.14  \\
            \hline
            TPSN         & 68.1$\pm$0.04  & 0.34$\pm$0.10 \\
            \hline
            TransAttUnet     &   73.5$\pm$0.05     & 0.32$\pm$0.09 \\
            \hline
            UA-MT       & 78.6$\pm$0.04  & 0.29$\pm$0.14  \\
            \hline
            BCP       & 80.1$\pm$0.038  & 0.256$\pm$0.133 \\
            \hline
            Ours      &   \textbf{83.6$\pm$0.04}  & \textbf{0.21$\pm$0.10} \\
            \hline
        \end{tabular}
        \label{tab1}
    \end{center}
\end{table}

\begin{figure}[htb]

\begin{minipage}[b]{1.\linewidth}
  \centering
  \centerline{\includegraphics[width=8.5cm]{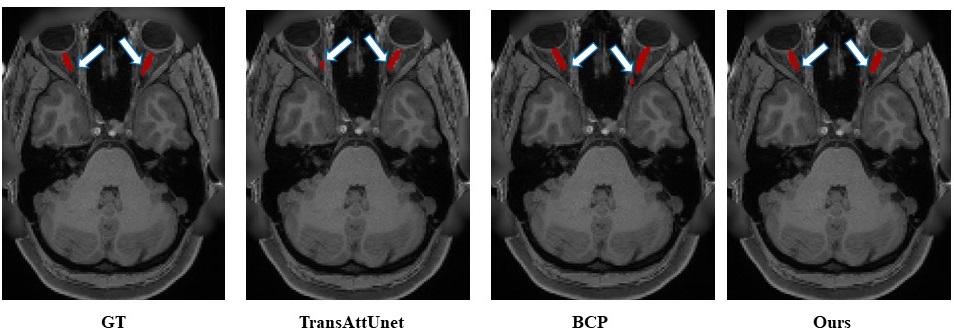}}
\end{minipage}
\caption{Visualization of VP segmentation results produced by different methods.}
\label{fig3}
\end{figure}

\begin{figure}[H]

\begin{minipage}[b]{1.\linewidth}
  \centering
  \centerline{\includegraphics[width=8.5cm]{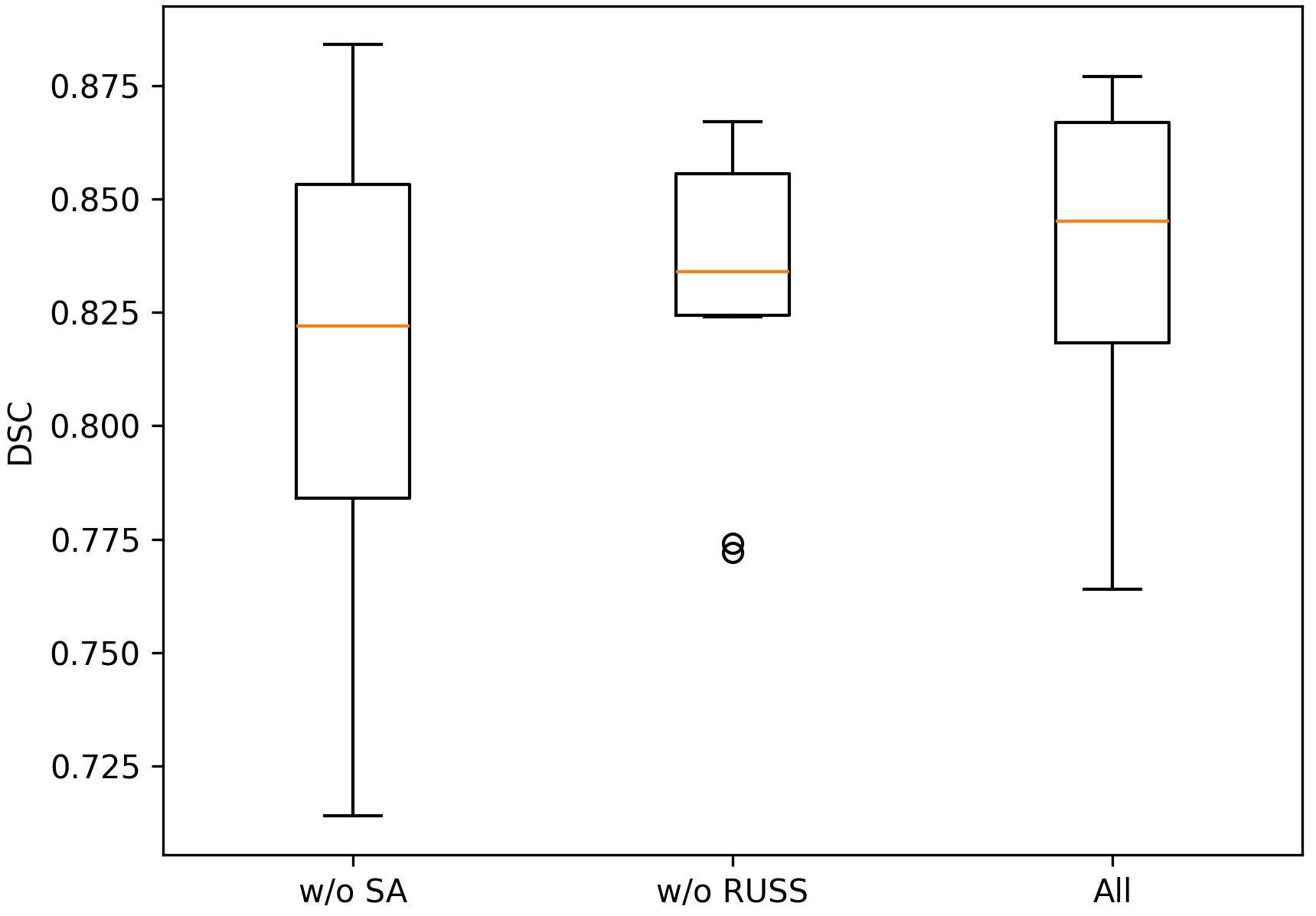}}
\end{minipage}
\caption{Effectiveness of each component on the HCP dataset. Here w/o RUSS means without the RUSS method, w/o SA means without the spatial attention, and All represents our proposed method with all components.}
\label{fig4}
\end{figure}

\textbf{Table \ref{tab1}} lists the quantitative segmentation results of different methods. Our proposed LESEN surpasses all other methods in terms of segmentation accuracy. LESEN achieves the highest Dice Similarity Coefficient (DSC) of 83.6\% and the lowest Average Symmetric Surface Distance (ASD) of 0.21 mm, validating the effectiveness of the proposed method for our task of VP segmentation.
Among the supervised methods, FuseNet achieved a DSC of 73.3\%, higher than the baseline methods (Baseline 1: 71.1\% and Baseline 2: 64.6\%), but lower than our proposed method. TPSN achieved a DSC of 68.1\%, lower than both FuseNet and our proposed method. TransAttUnet, a state-of-the-art supervised method utilizing attention mechanisms, achieved the highest DSC of 73.5\% among all the supervised methods. However, our proposed method surpassed TransAttUnet in terms of both DSC and ASD metrics, indicating the importance of exploiting information from unlabeled data.
For the semi-supervised methods, UA-MT achieved a DSC of 78.6\% and BCP achieved a DSC of 80.1\%. Although these methods performed well in leveraging unlabeled data, they were still surpassed by our proposed method, which achieved the highest DSC of 83.6\%. 
\textbf{Figure \ref{fig3}} plots the visualization results of different methods. Similar conclusions can be made that our proposed method achieves the best results.
Additionally, an ablation study was conducted to assess the effectiveness of different components of our method. As shown in \textbf{Figure \ref{fig4}}, the two components significantly impact our proposed approach.

\section{Conclusion}
In conclusion, this study presents LESEN, a label-efficient self-ensembling network, for multi-parametric MRI-based VP segmentation. The proposed LESEN addresses the challenge of limited labeled data by exploiting both labeled and unlabeled data. To achieve this, a self-ensembling mean teacher framework is built, allowing the student and teacher models to learn from each other. Additionally, a reliable unlabeled selection mechanism, RUSS, is introduced to further enhance the effectiveness of LESEN by selecting reliable unlabeled data for training. Experimental results on the HCP dataset demonstrate the superior performance of LESEN compared to state-of-the-art techniques. This advancement in multi-parametric MRI-based VP segmentation has significant implications for comprehensive analysis in clinical and research settings. Future research can explore the application of LESEN in other medical imaging tasks and further refine the reliable sample selection method to enhance its effectiveness. Overall, this work contributes to the field of multi-parametric MRI and paves the way for further advancements in visual pathway segmentation.

\section{Acknowledgments}
\label{sec:acknowledgments}

\begin{itemize} 
\item  This research was partly supported by the National Natural Science Foundation of China (62222118, U22A2040), Guangdong Provincial Key Laboratory of Artificial Intelligence in Medical Image Analysis and Application (2022B1212010011), Shenzhen Science and Technology Program (RCYX20210706092104034, \\ JCYJ20220531100213029), and Key Laboratory for Magnetic Resonance and Multimodality Imaging of Guangdong Province (2023B1212060052).
\item Data were provided in part by the Human Connectome Project, WU-Minn Consortium (Principal Investigators: David Van Essen and Kamil Ugurbil; 1U54MH091657) funded by the 16 NIH Institutes and Centers that support NIH Blueprint for Neuroscience Research; and by the McDonnell Center for Systems Neuroscience at Washington University.
\end{itemize}

\bibliographystyle{unsrtnat}
\bibliography{references}  






\end{document}